\documentclass[12pt]{iopart}
\usepackage{graphicx}

\def\ha{{1\over 2}}
\def\t{\tau}
\newcommand{\bl}{{\bf l}}

\newcommand{\bk}{{\bf k}}

\newcommand{\bV}{{\bf V}}

\newcommand{\bXp}{{\bf X}^\prime}
\newcommand{\bX}{{\bf X}}

\newcommand{\bR}{{\bf R}}
\newcommand{\bF}{{\bf F}}
\newcommand{\bZ}{{\bf Z}}
\newcommand{\bp}{{\bf p}}
\newcommand{\bP}{{\bf P}}

\newcommand{\bu}{{\bf u}}
\newcommand{\be}{{\bf \epsilon}}

\newcommand{\rmv}{{\rm v}}
\newcommand{\rmR}{{\rm R}}
\newcommand{\rmr}{{\rm r}}
\newcommand{\rms}{{\rm s}}
\newcommand{\rmx}{{\rm x}}
\newcommand{\rmt}{{\rm t}}

\newcommand{\rmb}{{\rm b}}
\newcommand{\w}{\omega}
\newcommand{\la}{\lambda}
\newcommand{\g}{\sf g}

\begin{document}
\title[Vacuum fluctuations and moving atoms/detectors:
From Casimir-Polder to Unruh]{Vacuum fluctuations and moving atoms/detectors:
From Casimir-Polder to Unruh Effect}

\author{B L Hu, A Roura and S Shresta}
\address{Department of Physics, University of Maryland, College Park, MD 20742}
\ead{hub@physics.umd.edu,sanjiv@physics.umd.edu,roura@physics.umd.edu}

\begin{abstract}

In this note we report on some new results~\cite{SHP} on
corrections to the Casimir-Polder~\cite{caspol} retardation
force due to atomic motion and present a preliminary
(unpublished) critique on one recently proposed cavity QED
detection scheme of Unruh effect~\cite{Unr76}. These two
well-known effects arise from the interaction between a moving
atom or detector with a quantum field under some boundary
conditions introduced by a conducting mirror/cavity or dielectric
wall.

The Casimir-Polder force is a retardation force on the atom due
to the dressing of the atomic ground state by the vacuum
electromagnetic field in the presence of a conducting mirror or
dielectric wall. We have recently provided an improved
calculation by treating the mutual influence of the atom and the
(constrained) field in a self-consistent way.  For an atom moving
adiabatically, perpendicular to a mirror, our result finds a
coherent retardation correction up to twice the stationary value.

Unruh effect refers loosely to the fact that a uniformly
accelerated detector feels hot. Two prior schemes have been
proposed for the detection of `Unruh radiation', based on charged
particles in linear accelerators and storage rings. Here we are
interested in a third scheme proposed recently by Scully~{\it et
al}~\cite{Scully03} involving the injection of accelerated atoms
into a microwave or optical cavity. We analyze two main factors
instrumental to the purported success in this scheme, the cavity
factor and the sudden switch-on factor.   We conclude that the
effects engendered from these factors are unrelated to the Unruh
effect.

\end{abstract}
\maketitle

\section{Introduction}

In this short note we report on some new results and present some
thoughts on two well-known effects arising from the interaction
between a moving atom or detector with a quantum field. The
Casimir-Polder force \cite{caspol} is a retardation force on the
atom due to the dressing of the atomic ground state by the vacuum
electromagnetic field in the presence of a boundary (e.g., a
mirror or a cavity). We set forth to improve on existing
calculations by treating the mutual influence of the atom and the
(constrained) field in a self-consistent way. This is in order to
maximally preserve the coherence of the combined system, which is
a highly desirable if not critically demanded criteria for quantum
computer designs. Since a self-consistent treatment requires
backreaction considerations, the atom's motion should be included
to account for the full effect of the field on the atom. For an
atom moving adiabatically, perpendicular to a mirror, our recent
result finds a coherent retardation correction up to twice the
stationary value~\cite{SHP}.

The Unruh effect \cite{Unr76} described colloquially states that
a uniformly accelerated detector~(UAD) feels hot at the Unruh
temperature. There are at least three classes of detection
schemes proposed. One based on charged particles in linear
accelerators purports to measure at a distance the radiation
emitted from the UAD~\cite{Chen}. Researchers have commonly
agreed that a uniformly accelerated detector does not emit
radiation. (See, e.g.,~\cite{Grove,RSG,Unr92,MPB,RHA,RHK,HRCapri}
and references therein). If there were emitted radiation it would
have to be from nonuniform acceleration and treated by
nonequilibrium quantum field theory concepts and
techniques~\cite{HJCapri,JH1}. But these are not the Unruh
effect. Detection of Unruh effect in storage rings has earlier
been proposed~\cite{BelLen}. Even though acceleration in the
circular case shares some features with linear acceleration which
is what Unruh effect entails, there are basic differences between
these two cases. For one, in circular motion acceleration comes
from changes in the direction, not in the speed. Controversies of
its nature and viability remain~\cite{QABP}.

Recently a detection scheme has been proposed by Scully {\it et
al}~\cite{Scully03} involving the injection of accelerating atoms
into a microwave or optical cavity. It is this scheme which we
wish to discuss here. We first identify the main working parts in
this scheme and then analyze their effects and relevance to the
Unruh effect. The main factors we shall consider here are a) the
effect of cavity fields on the accelerating atom in the emission
of photons and b) the effect of sudden switching on of the atom -
field interaction from the injection of atoms into the cavity.
For the cavity factor we find that a thermal distribution of the
photons in the cavity should not be identified as a manifestation
of the Unruh effect. As for the sudden switch-on effect, photons
are produced from the nonadiabatic amplification of vacuum
fluctuations (similar to cosmological particle
creation~\cite{BirDav}), but not from the atom's uniform
acceleration (analogous to Hawking radiance in black
holes~\cite{Haw74}).

Note that in both effects there are three dynamical variables:
the internal degrees of freedom (dof) of a detector -- a two
level system, the external degrees of freedom -- the center of
mass~(COM) of the detector/atom in motion, and the quantum field
(scalar or electromagnetic) under some constrained condition
imposed by the presence of the mirror or cavity. (One can easily
extend this class of problems to  cases where the mirror is in
motion \cite{FulDav,RHK} giving rise to  dynamical Casimir
effects \cite{dynCas} and analogs of Hawking effect \cite{Haw74}
in quantum black holes.) In Casimir-Polder effect it is the
external dof -- the COM -- of the atom interacting with the field
modes altered by the mirror which are of interest, whereas in the
Unruh effect, as the trajectory is prescribed, it is the internal
dof of the detector/atom and the field correlations which are of
interest. A more obvious difference is, of course, the atom in
Casimir-Polder undergoes nonrelativistic motion, whereas it is
the uniform acceleration engendering an event horizon which gives
the distinct thermal feature in the Unruh radiation. But
ultimately these are only different regimes and manifestations of
the same set of problems related to the kinematic and boundary
effects on the quantum vacuum. We deliberately bring these two
effects into the same discussion so as to stimulate stronger cross
fertilization between these two apparently disparate fields,
namely, quantum/atom optics and quantum field theory in curved
spacetime \cite{BirDav}. It is the belief of one of us that many
fundamental effects in the latter discovered theoretically in the
70's may find verifications in the experimental schemes of the
former. What is more interesting is that there is plenty of room
in their intersection for asking new sets of questions pertaining
to the (nonequilibrium) statistical mechanical properties of the
quantum vacuum as influenced by moving atoms, detectors, mirrors
or background fields such as vacuum viscosity due to particle
creation \cite{Zel70,HuVacVis,KME}, vacuum friction due to atom
motion \cite{Pendry,Kardar}, quantum noise under exponential
red-shifting \cite{Dalian,HRSV}, field correlations in spacetimes
with event horizons \cite{HRCapri,GHP} and entanglement and
teleportation with non-inertial observers \cite{Milburn,RHA}.

This paper is evenly divided into two parts: Sections~2,~3 deal
with the correction to the Casimir-Polder effect due to atom
motion, Sec.~4 summarizes the main features of Unruh effect,
Sec.~5 discusses the scheme of Scully {\it et al} for the detection of
Unruh effect by accelerating atoms in a cavity.

  \vskip .5 cm \centerline {\bf Part I:
Casimir-Polder Effect} \vskip .5 cm

Recent experiments~\cite{suk1,lan1} have reported on the
measurement of the Casimir-Polder~(C-P) retardation
force~\cite{caspol}, which is a retardation force arising from the
quantum modification of the electrostatic attraction of a
polarizable atom to its image in the wall.
\begin{figure}[ht]
\begin{center}
\includegraphics[width=4cm]{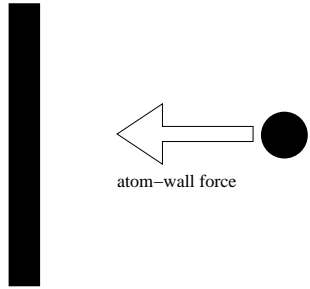}
\caption{The force between an atom and a wall is a combination of
the electrostatic force, the Casimir-Polder retardation force, and
the correction to the retardation as described here.}
\end{center}
\end{figure}
In its usual interpretation the retardation force arises from the
dressing of
the atomic ground state by the vacuum electromagnetic field~(EMF)
in the presence of a boundary. With this interaction, the ground
state of the combined atom-EMF system is no longer a product of
the separate free ground states, but is instead an entangled
atom-EMF state. The retardation force originates from the quantum
correlation between two interacting systems (or, if one of them
is of less concern in the observation, one system S -- such as
the atom, and its environment E -- the quantum field). Another
manifestation is colored noise from the coarse-grained quantum
fields. To follow the quantum correlations between S and E, we
need to adopt a method which can maximally preserve the quantum
coherence in treating the S and E interaction consistently.

The state-of-the-art derivation~\cite{Milonni,barton1,Powers} of
the Casimir-Polder retardation force is by calculating the
gradient of a spatially dependent dressed ground state. Here we
report on a rederivation~\cite{SHP} of the C-P retardation force
given in terms of the recoil associated with the emission and
reabsorption of virtual photons. While the gradient calculation
of the force assumes a stationary atom, our method treats an atom
that moves adiabatically.

Our result in the stationary atom limit is in exact agreement with
the Casimir-Polder force. In the case of an adiabatically moving
atom, we find a coherent retardation correction up to twice the
stationary value. The additional correction due to atomic motion
can be understood 
as indicating that the dressed ground states for stationary and
moving atoms are not the same, the difference arising from the
Doppler shift of the EMF modes with respect to the conducting
wall. That is, a moving atom is in a Doppler shifted vacuum, so
its dressed ground state is altered from the stationary one. In
addition, what is usually pictured physically in terms of a
gradient force is explained  here in terms of the recoil
associated with emission and absorption of virtual photons.
\footnote{After our work \cite{SHP} was published Professors Paul
Davies and Gerard Milburn brought to our attention Pendry's work
\cite{Pendry} on vacuum friction due to atoms moving near an
imperfectly conducting surface.  Pendry's interpretation of the
force in his case is similar to ours here for the Casimir-Polder
retardation force arising from the recoil of Doppler shifted
emitted and absorbed virtual photons.  Note atoms are moving
parallel in Pendry's case whereas in ours they are perpendicular
to the wall.} This provides a quantum field theory interpretation
of the dipole force~\cite{CCT1}

This work is relevant to applications in which atoms are trapped
on the order of a resonant atomic wavelength near a surface.
Examples include evanescent wave gravito-optical~\cite{evan},
microlens array~\cite{birkl1}, and magnetic chip
trapping~\cite{chip}. Recent experiments have demonstrated the
measurable effects of retardation on atomic motion near a
surface~\cite{suk1,lan1}. Those effects will become more
important as such applications become more refined. That is
especially true when precision control over the motion of atoms
is needed, such as the implementation of two-qubit gates in AMO
quantum computer designs.

\section{Approach}
We assume an atom placed near a conducting wall is initially in a
factorizable state with the EMF vacuum. The combined system
evolves according under the minimal coupling QED Hamiltonian in
the dipole approximation, but without the rotating wave
approximation
\begin{eqnarray}
\label{hamiltonian1} H =\frac{{\bf P}^2}{2M}
+\hbar\w_0 S_+ S_- +\hbar\sum_\bk \w_\bk \rmb_\bk^\dagger \rmb_\bk
+H_{I} = H_0 + H_I.
\end{eqnarray}
The operators $S_{\pm}$ are the up and down operators of the
atomic qubit and $\w_0$ is the atomic transition frequency. The
operators  $\rmb_\bk$ and $\rmb_\bk^\dagger$ are the EMF mode
annihilation and creation operators, and $\w_\bk$ are the
frequencies of the EMF modes. The interaction Hamiltonian is made
up of two parts $ H_{I}= H_{I1} + H_{I2} $,
\begin{eqnarray}
\label{interaction1} H_{I1} &=& \hbar\sum_{\bk e} \frac{\g}{\sqrt{\w_\bk}} [\bp_{eg} S_+ +\bp_{ge} S_-]\cdot[\bu_\bk \rmb_\bk +\bu_\bk^\dagger \rmb_\bk^\dagger] \\
\label{interaction2} H_{I2} &=& \hbar\sum_{\bk\bl} \frac{\la^2}{\sqrt{\w_\bk \w_\bl}} [\bu_\bk\cdot\bu_\bl \rmb_\bk \rmb_\bl +\bu_\bk^\dagger\cdot\bu_\bl(\delta_{\bk\bl} +2\rmb_\bk^\dagger \rmb_\bl ) +\bu_\bk^\dagger\cdot\bu_\bl^\dagger \rmb_\bk^\dagger \rmb_\bl^\dagger].
\end{eqnarray}
The vector $\bp_{eg} = \langle e| \bp | g\rangle = -\rmi m\w_0
\langle e| {\bf r} | g\rangle$ is the dipole transition matrix
element between the ground (g) and excited (e) states. The vectors
$\bu_\bk(\bX) = \hat{\be}_\bk f_\bk(\bX)$ contain the photon
polarization vectors $\hat{\be}_\bk$ and the spatial mode
functions $f_\bk(\bX)$. The coupling constants are $\g =
-\sqrt{8\pi^2\alpha c/m^2}$ and $\la=\sqrt{4\pi^2\hbar\alpha
c/m}$, with $\alpha$ being the fine structure constant. In the
presence of a conducting plane boundary, the spatial mode
functions of the EMF which satisfy the imposed boundary
conditions are the TE and TM polarization modes~\cite{Milonni82},
\begin{eqnarray}
\label{mode functions 1} \bu_{\bk 1}(\bX) &=& \sqrt{\frac{2}{L^3}}\hat{\bk}_\|\times\hat{\bZ} \sin(k_Z Z) e^{\rmi\bk_\|\cdot\bX} \\
\label{mode functions 2}\bu_{\bk 2}(\bX) &=& \sqrt{\frac{2}{L^3}}\frac{1}{k}[k_\|\hat{\bZ} \cos(k_Z Z) -\rmi k_Z \hat{\bk}_\| \sin(k_Z Z)] e^{\rmi\bk_\|\cdot\bX},
\end{eqnarray}
and their complex conjugates.

The transition amplitude for the atom to move  from some initial
to final position, while remaining in the atomic ground state
with the EMF in the vacuum is,
\begin{equation}
K[\bX_f; \rmt+\tau, \bX_i; \rmt]  = \mbox{}_o\langle \bX_f;
\rmt+\tau | \exp[-\frac{\rmi}{\hbar}\int_{\rmt}^{\rmt+\tau}
H(\rms) \rmd\rms] | \bX_i; \rmt \rangle_o
\end{equation}
for which the expectation value of the momentum operator is
derived.
\begin{eqnarray}\fl
\label{momentum expectation} \langle \hat{\bP} \rangle
(\rmt+\tau)  = \frac{\hbar}{N} \int \frac{\rmd\bP_f}{(2\pi)^3} \\
\nonumber \times\mbox{ }\bP_f\int\rmd\bX_i \rmd\bXp_i \mbox{
}K[\bP_f;\rmt+\tau|\bX_i; \rmt] \mbox{ }\Psi(\bX_i)
\Psi^*(\bX^\prime_i) \mbox{ }K^*[\bP_f;\rmt+\tau|\bXp_i; \rmt],
\end{eqnarray}
with the normalization factor
\begin{equation}\fl
N = \int \frac{\rmd\bP_f}{(2\pi)^3} \int\rmd\bX_i \rmd\bXp_i \mbox{ }K[\bP_f;\rmt+\tau|\bX_i; \rmt] \mbox{ }\Psi(\bX_i) \Psi^*(\bX^\prime_i) \mbox{ }K^*[\bP_f;\rmt+\tau|\bXp_i; \rmt].
\end{equation}
The transition amplitude is evaluated as a path integral in
which  Grassmannian and bosonic coherent states are used to label
the atomic and EMF degrees of freedom,
respectively~\cite{ohnuki,cahill,perelomov,gilmore}. The position
and momentum basis are used for the atom's center of mass degree
of freedom. The major approximation applied is a resummed 2nd
order vertex approximation. The 2nd order vertex approximation
preserves the coherence of the combined system  at long and short
times, as it is a partial resummation of all orders of the
coupling. The resulting transition amplitude is
\begin{eqnarray}\fl
\label{classical transition amplitude}
K[\bX_f; \rmt+\tau, \bX_i; \rmt] =
\bigg( \frac{M}{2\pi\rmi\hbar\tau}\bigg )^{3/2} \exp\bigg\{ \rmi
\int_{\rmt}^{\rmt+\tau} \bigg[\frac{M\dot{\bX_c^0}^2}{2\hbar}  \nonumber\\
+\rmi\bp_z^2 \int_{\rmt}^\rms \rmd\rmr \sum_\bk \frac{\g^2}{\w_\bk}
\mbox{ }\rme^{-\rmi(\w_\bk +\w_0)(\rms-\rmr)}\bu_\bk (\bX_c^0(\rms))
\cdot\bu_\bk^*(\bX_c^0(\rmr)) \nonumber\\
-\sum_\bk \frac{\la^2}{\w_\bk} \bu_\bk^*(\bX_c^0(\rms)) \cdot\bu_\bk(\bX_c^0(\rms))
+O(\rme^4/M)\bigg]\rmd\rms \bigg\}.
\end{eqnarray}
Assuming the position wavefunction of the atom is a Gaussian
wavefunction centered  at $(\bR, \bP_0)$ with the standard
deviations $(\sigma, 1/\sigma)$, the force on the atom, which is
the time derivative of the momentum expectation value is found to
be
\begin{eqnarray}\fl
\label{general force} \bF_c(\rmR,\rmv,\rmt+\tau)
= -\frac{2\pi\rmi\alpha_0 \hbar\w_0^2}{L^3}
\sum_\bk \frac{\bk_z \cos^2\theta}{\w_\bk}
\mbox{ }\rme^{-2\rmi\bk_z \cdot(\bR +\bV\tau)} \nonumber\\
+\frac{\pi\alpha_0 \hbar\w_0^3}{L^3}
\sum_\bk \frac{\bk_z \cos^2\theta}{\w_\bk}
\int_{\rmt}^{\rmt+\tau} \rmd\rms \mbox{ }\rme^{-\rmi\bk_z \cdot(2\bR +\bV (\tau+\rms-\rmt))} \\ \nonumber
\mbox{ }\mbox{ }\mbox{ }\mbox{ }\mbox{ }\mbox{ }\mbox{ }\mbox{ }\mbox{ }\mbox{ }
\times\bigg[ \rme^{-\rmi(\w_k +\w_0)(\rmt+\tau-\rms)} -\rme^{\rmi(\w_k +\w_0)(\rmt+\tau-\rms)}
 \bigg],
\end{eqnarray}
where $\alpha_0$ is the static ground state polarizability. The
mass of the atom has been taken to infinity and its extension to
a point, while  finite terms are retained to their effect on the
dynamics.

\section{Results and Interpretation}
\subsection{stationary atom}
The total force on a stationary atom is given by the sum of the
correction force, while setting $\rmv =0$,  and the electrostatic
force. The stationary atom force exhibits a transient behavior
when the atom first "sees" itself in the wall. Then, on a
timescale of several atom-wall round trip light travel times it
asymptotes to the following constant steady state value
\begin{eqnarray}
\label{stationary force}\bF_{sa}(\rmR, \tau>>2\rmR/c)  = \hat{\bf
e}_z \frac{\alpha_o\hbar\w_0^2}{8\pi } \bigg(\frac{\rmd}{\rmd
\rmR}\bigg)^3 \frac{1}{\rmR} \int_0^\infty \frac{\rmd\rmx}{\rmx^2
+\w_0^2} \mbox{ }\rme^{-2\rmR \rmx/c}.
\end{eqnarray}
From Eq.~(\ref{stationary force})  the potential which a
stationary atom feels is easily found to be
\begin{eqnarray}
\label{stationary potential}U_{sa}(\rmR)
= -\frac{\alpha_o\hbar\w_0^2}{8\pi } \bigg(\frac{\rmd}{\rmd \rmR}\bigg)^2 \frac{1}{\rmR}
\int_0^\infty \frac{\rmd\rmx}{\rmx^2 +\w_0^2} \mbox{ }\rme^{-2\rmR \rmx/c},
\end{eqnarray}
with asymptotic limits
\begin{equation} \label{stationary limits}
\begin{array}{lr}
U_{sa}(\rmR) \rightarrow -\frac{\alpha_o\hbar\w_0}{8 }
\frac{1}{\rmR^3} & \mbox{ for } \rmR << \frac{c}{\w_0} \\
U_{sa}(\rmR) \rightarrow -\frac{3\alpha_o\hbar c}{8\pi }
\frac{1}{\rmR^4} & \mbox{ for } \rmR >> \frac{c}{\w_0}
\end{array},
\end{equation}
which exactly reproduces the results from the energy gradient
approaches~\cite{caspol}.

\subsection{moving atom}
The retardation force for a moving atom can be determined  from
Eq.~(\ref{general force}) by applying a separation of short time
scale dynamics from long time scale dynamics analogous to
standard methods for determining the dipole force on an atom in a
laser beam~\cite{CCT1}. There, assuming that the atom's position
is constant on short timescales, the optical Bloch equations are
solved for the steady state values of the internal state density
matrix elements. On long time-scales the matrix elements are
replaced by their steady state values and put into the Heisenberg
equation of motion for the atomic COM momentum. Such a procedure
is justified when the internal and external dynamics evolve on
vastly different timescales. The analogous separation here will
be of the short timescale describing the self-dressing of the
atom-EMF system and the long timescale describing the motion of
the atom.

After evaluating adiabatically and combining the retardation correction force with the electrostatic force, the atom-wall force is found to be
\begin{eqnarray}\fl
\label{adiabatic force}\bF_{am}(\rmR) = \hat{\bf e}_z \frac{\alpha_o\hbar\w_0^2}{8\pi } \bigg(\frac{\rmd}{\rmd \rmR}\bigg)^3 \frac{1}{\rmR} \int_0^\infty \frac{\rmd x}{x^2 +\w_0^2} \mbox{ }\rme^{-2\rmR x/c} \\ \nonumber
-\hat{\bf e}_z \frac{\alpha_o\hbar\w_0^2}{4\pi } \bigg(\frac{\rmd}{\rmd\rmr}\bigg)^3 \int_0^\infty \frac{\rmd k}{k c+\w_0} \frac{\sin(2k\rmr)}{2k\rmr}\Bigg|^\rmR_{\rmR_0}.
\end{eqnarray}
The first term is the stationary atom-wall force and the second term is a residual force which pulls the atom back to its original point of release. The force can easily be turned into the potential which the atom feels:
\begin{eqnarray}\fl
\label{adiabatic potential}U_{am}(\rmR) = -\frac{\alpha_o\hbar\w_0^2}{8\pi } \bigg(\frac{\rmd}{\rmd \rmR}\bigg)^2 \frac{1}{\rmR} \int_0^\infty \frac{\rmd x}{x^2 +\w_0^2} \mbox{ }\rme^{-2\rmR x/c} \\ \nonumber
+ \frac{\alpha_o\hbar\w_0^2}{4\pi } \bigg(\frac{\rmd}{\rmd\rmr}\bigg)^2 \int_0^\infty \frac{\rmd k}{k c+\w_0} \frac{\sin(2k\rmr)}{2k\rmr}\Bigg|^\rmR_{\rmR_0}.
\end{eqnarray}
Since the first term in the potential is the stationary atom-wall potential, in the regions near and far from the wall it will have the expected inverse powers of distance dependence, as shown in Eq.~(\ref{stationary limits}). The second term is the residual potential due to the motion.

\subsection{Conclusion} In the traditional (energy gradient)
approach, one interprets the force between a polarizable atom and
a wall as arising from the Lamb shift in the atomic ground state
energy. Spatial variation of the ground state energy is expected
to generate a force which pushes the atom perpendicular to the wall,
but the mechanism for such a force is not given explicitly.
Our approach provides an interpretation of how a net force arises
from the emission-reabsorption processes in the presence of a
boundary. It goes beyond Lamb shift calculations in that it
incorporates the effect of slow atomic motion.

The extra correction from our coherent QED calculation makes a
verifiable prediction.  The alteration of the force has its best
chance of being measured in experiments involving cold atoms
bouncing off the evanescent field of a laser beam totally
internally reflected in a crystal. In those experiments the laser
is blue detuned, which imposes a repulsive potential to counter
the attractive potential of the wall thus creating a barrier for
cold atoms moving toward the crystal to bounce against. As the
intensity of the evanescent laser field is lowered the height of
the barrier is lowered. At some threshold value the barrier
height will fall below the classical tunnelling height and no
atoms will be reflected. The van~der~Waals, Casimir-Polder, and
our coherent QED (corrected Casimir-Polder) forces all give
different predictions for that threshold laser intensity. The
calculations done here are for a perfect conductor, not a
dielectric boundary, so the modifications predicted here should
not be applied directly to the case of a dielectric boundary.
However, a general statement can be made that a coherent QED
correction will cause a lowered prediction for the threshold
laser power, since it will tend to decrease the atom-wall
attraction. If one naively applies a dielectric factor to our
result for the conducting plate to compensate for the difference,
the present prediction for the threshold energy in units of the
natural line width (14.8 $\Gamma$) is closer to the measured
value (14.9$\pm1.5\mbox{ }\Gamma$), compared to the previously
predicted value of (15.3 $\Gamma$)~\cite{lan1}.

\vskip .5cm \centerline{\bf Part II. Detection of Unruh Effect by
Accelerating Atoms in Cavity } \vskip .5cm

\section{Unruh Effect: Main Features and Common Misunderstandings}

Unruh effect attests that a particle detector following a
uniformly accelerated trajectory~\footnote{In special relativity,
an object is considered to be uniformly accelerated if the
acceleration always remains the same as measured at each instant
of time by the observers in the inertial frame where the object
is at rest at that time.} in Minkowski spacetime perceives the
vacuum of the field as a thermal bath. It is a purely quantum
mechanical effect since it originates from the vacuum quantum
fluctuations of the field and  is nonexistent for a classical
field.  We define a detector as any quantum system with some
internal degree of freedom which couples to a quantum environment.
Unruh used a monopole detector interacting  with a scalar field in
his exposition. It could also be a two-level atom interacting
with an electromagnetic field.  In this section we briefly
discuss several fundamental features of Unruh effect whose lack of
understanding has caused some confusion in some experimental
proposals for its detection. We then focus on the Scully {\it et
al} scheme involving accelerating atoms in a cavity field.



\subsection{Interaction between a Moving Detector and a Quantum Field }

We model  the moving  detector by a harmonic oscillator locally
coupled to a massless real scalar field $\phi$ in $1+1$
dimensions and follow the discussions of~\cite{RHA,RHK}. The total
action $S[Q,\varphi] = S_\mathrm{osc}[Q] + S_\mathrm{f}[\varphi]
+ S_\mathrm{int}[Q,\phi]$ is the sum of three contributions: The
action for a free harmonic oscillator of mass $M$ and frequency
$\Omega$ is
\begin{equation}
S_\mathrm{osc}[Q] = \int \rmd\tau \left[
\frac{1}{2}M\dot{Q}^{2}(\tau) - \frac{1}{2} M \Omega
^{2}Q^{2}(\tau) \right],
\end{equation}
where $Q(\tau)$ is the internal coordinate of the oscillator which
follows a trajectory $x^{\mu} = (t (\tau), x(\tau))$ parametrized
by its proper time $\tau$. The action for the free massless field
$\phi$ is
\begin{equation}
S_\mathrm{f}[\phi] = \int \rmd t \rmd x
\left[\frac{1}{2}(\partial_t \phi(x^\mu))^2  - \frac{1}{2}
(\partial_x \phi (x^\mu) )^2 \right].
\end{equation}
The interaction action is
\begin{equation}
S_\mathrm{int}[Q,\phi] = - \lambda \int \rmd\tau {Q}(\tau)
\frac{d\phi}{d \tau} (x^\mu(\tau)),
\end{equation}
where $\lambda$ is the coupling constant between the field and the
detector, represented by the harmonic oscillator. The detector
couples locally to the field at the spacetime points along its
trajectory. (For details in the transform between $t$ and $\tau$
in the integration limits, see~\cite{RHA}.)

The dynamics of the detector (our system S) can be studied by
using the influence functional formalism~\cite{if} for quantum
open systems,  viewing the field as its environment (E). The
state of the detector is described by the reduced density matrix
$\rho_\mathrm{r} (Q_i,Q'_i,t_i)$. At the initial time $t_i$
assume that the field and the detector are uncorrelated. The
reduced density matrix for the detector at a later time $t_f$ is
given formally by
\begin{equation}
\fl\rho_\mathrm{r} (Q_f,Q'_f,t_f) = \int \rmd Q_i \rmd Q'_i
\int\limits_{Q_i} ^{Q_f} \mathcal{D}Q \int\limits_{Q'_i}^{Q'_f}
\mathcal{D}Q' \rme^{\rmi(S[Q]-S[Q']+S_{IF}[Q,Q']) / \hbar} \rho_r
(Q_i,Q'_i,t_i).
\end{equation}
The influence action $S_\mathrm{IF}$ has the following expression:
\begin{eqnarray}
\fl S_\mathrm{IF}[Q,Q'] = -2 \int_{\tau_i}^{\tau_f} \rmd\tau
\int_{\tau_i}^{\tau} \rmd\tau' \Delta (\tau) D(\tau,\tau')
\Sigma(\tau') \\ \nonumber
+ \frac{\rmi}{2} \int_{\tau_i}^{\tau_f} \rmd\tau
\int_{\tau_i}^{\tau_f} \rmd\tau' \Delta (\tau) N(\tau,\tau')
\Delta (\tau'),
\end{eqnarray}
where $\Delta(\tau) \equiv Q'(\t) - Q(\t), \;\;\; \Sigma (\t)
\equiv \ha [Q'(\t) +Q (\t)]$ and the dissipation and noise
kernels are given respectively by (note these simple expressions
are valid only for linear S-E coupling, see,~\cite{RHK}):
\begin{eqnarray}
D(\tau,\tau') = \frac{\rmi}{2} \left\langle \left[
 \frac{d\hat{\phi}(\tau)}{d \tau},
 \frac{d\hat{\phi}(\tau')}{d\tau'} \right] \right\rangle , \\
N(\tau,\tau') = \frac{1}{2} \left\langle \left\{
 \frac{d\hat{\phi}(\tau)}{d\tau},
 \frac{d\hat{\phi}(\tau')}{d\tau'}\right\} \right\rangle
...
\end{eqnarray}
Here $[, ]$ and $\{, \}$ denote the commutator and anti-commutator
respectively, and the expectation values are taken with respect to
the initial state of the field, assumed to be the Minkowski
vacuum.

A uniformly accelerated trajectory for the detector is
characterized by
\begin{equation}
t(\tau) = \alpha^{-1} \sinh \alpha \tau ,  \;\;\; x(\tau) = c
\alpha^{-1} \cosh \alpha \tau , \label{accel}
\end{equation}
where $c$ is the speed of light and $\alpha c$ is a constant that
corresponds to the acceleration measured by the instantaneously
comoving inertial observer at each instant of time. When such a
trajectory is considered, both the dissipation and noise kernels
are stationary, i.e. $D(\tau,\tau') = D(\tau - \tau')$ and
$N(\tau,\tau') = N(\tau - \tau')$. This follows from the fact
that both the uniformly accelerated trajectory given by
equations~(\ref{accel}) and the Minkowski vacuum state are
invariant under Lorentz transformations. (A Lorentz
transformation simply corresponds to a constant shift of $\tau$
in equations~(\ref{accel}.) What is more, while the dissipation
kernel remains the same as in the case of an inertial detector,
the noise kernel is equivalent to the result that one would
obtain for an inertial detector if a thermal state rather than
the vacuum were chosen as the initial state for the
field.~\footnote{ Due to the singular behavior of the dissipation kernel,
in most cases it is necessary to consider smeared trajectories
for the detector~\cite{JH1}. In some of those cases the limit in
which the size of the smearing goes to zero can be taken after
renormalizing the frequency of the oscillator-detector. This
smearing can be introduced in a way that respects  Lorentz
invariance. Simply multiply the right-hand side of
equations~(\ref{accel}) by a factor $(1 + c^{-2} a d)$, which
gives rise to trajectories at a constant physical distance $d$
from the original trajectory as measured by the accelerated
observers, and integrate over some smearing function $f(d)$
centered at $d=0$, such as $\exp(-x^2/\sigma^2)$, where $\sigma$
is the characteristic smearing scale in this case.}

We have modeled our detector by a harmonic oscillator, but one
could alternatively have considered a two-level system with
natural frequency $\omega_0$  described by the free action
\begin{equation}
S_\mathrm{2LS}[S_{+},S_{-}] = \int \rmd\tau\, \hbar \omega_0
S_{+}(\tau) S_{-}(\tau) ,
\end{equation}
and an interaction term of the form
\begin{equation}
S_\mathrm{int}[S_{+},S_{-};\phi] = \lambda \int \rmd\tau \left(
S_{+} (\tau) + S_{-} (\tau) \right) \phi (x^\mu(\tau)),
\end{equation}
where $S_{+}$ and $S_{-}$ are the Grassmann variables that, when
quantizing, give rise to the up and down operators for the
two-level system, and $\hbar\omega_0$ is the energy difference
between the excited state and the ground state.

\subsection{Stationarity, Lorentz invariance and absence of real
emitted radiation}

The behavior of the accelerated detector has the following
characteristic features (see~\cite{RHA,RHK,JH1},~\cite{MPB,MP1}
and related work quoted therein). During an initial transient
time, owing to the term involving the dissipation kernel, all
the information on the initial state of the detector is dispersed
into the field, while the detector thermalizes and reaches an
equilibrium condition with the vacuum fluctuations of the field,
perceived as a thermal bath. After this transient time, the
detector remains stationary, implying that the combined system of
detector plus field is invariant under the Lorentz
transformations. Making use of that symmetry one can easily argue
that the energy flux of the field must vanish.

Indeed, for the model above, the self-consistent dynamics of the
detector and the field, including their interaction, can be
solved exactly. One finds that indeed the expectation value of the
stress tensor operator for the field vanishes. However, even
though there is no flux of energy emitted from the detector, the
vacuum polarization is altered due to the emission and
reabsorption of virtual particles by the accelerated detector.
This can be verified by computing the two-point quantum
correlation function of the field (which is still Lorentz
invariant) and comparing to the case without detector (the
correlation function for the free field in the vacuum
state)~\cite{HRCapri}.

There are many subtle points underlying the absence of emitted
radiation or energy flux in the field  from a UAD. Unruh effect
is predicated upon the condition that the uniform acceleration
goes on for an indefinite amount of time,  which requires an agent
doing work on it. If  the detector has been accelerating for a
finite duration~\cite{RHK} or if the interaction lasts for a
finite period of time but switched on adiabatically~\cite{MP1}
there will be radiation emitted. It also makes a difference
whether the backreaction of the field on the detector is
accounted for. For a preliminary consideration including
backreaction, see~\cite{parentani95}. For a fully self-consistent
backreaction calculation involving moving charges,
see,~\cite{JH1}. We hope to return to these issues in a future
investigation~\cite{HJR}.

\section{Accelerated atoms in optical or microwave cavities}

We now describe a scheme recently proposed by Scully {\it et
al}~\cite{Scully03} to detect phenomena related to the Unruh effect
based on accelerating atoms inside microwave (or optical)
cavities. We will only highlight the main features and discuss
their relevance to Unruh effect.

\subsection{Relevant factors in the detection scheme of Scully {\it et al}}

In their theoretical analysis they consider a model consisting of
a two-level atom and a single electromagnetic mode inside the
cavity. They evaluated the change of the reduced density matrix
for the electromagnetic mode when an atom is injected in a cavity
with the electromagnetic field in its ground state by working in
the interaction picture and treating the interaction term
perturbatively.  They gave an expression for an atom which is
injected at the proper time $\tau_\mathrm{i}$ with zero initial
velocity to quadratic order (the lowest nontrivial order)as
\begin{equation}
\fl \delta \hat{\rho}_\mathrm{i} = - \frac{1}{\hbar^2}
\int_{\tau_\mathrm{i}} ^{\tau_\mathrm{i}+ T}
\int_{\tau_\mathrm{i}}^{\tau_\mathrm{i}+\tau'}
 \Tr_\mathrm{2LA} \left[\hat{H}_\mathrm{int}(\tau'),
 \left[\hat{H}_\mathrm{int}(\tau''),
\hat{\rho}_\mathrm{2LA}(\tau_\mathrm{i}) \otimes
\hat{\rho}_\mathrm{em}(t(\tau_\mathrm{i})) \right] \right]
\rmd\tau' \rmd\tau''  \label{single} ,
\end{equation}
where $T$ is the proper time of flight inside the cavity as
measured by the atom and $Tr_\mathrm{2LA}$ denotes the partial
trace with respect to the states of the two-level atom. Assuming
that the dipole approximation is valid in the instantaneous rest
frame~(IRF) for the atom, they write the  atom-field interaction
Hamiltonian $\hat{H}_\mathrm{int}$ as
\begin{equation}
\fl \hat{H}_\mathrm{int}(\tau) = \hbar \lambda(\tau)
\left(\hat{a}_k \rme^{-\rmi \nu t(\tau) + \rmi k_x x(\tau)} +
\hat{a}_k^\dagger \rme^{\rmi \nu t(\tau) - ik_x x(\tau)}\right)
\left( \hat{S}_{+} \rme^{-\rmi\omega\tau} + \hat{S}_{-}
\rme^{\rmi\omega\tau}\right) ,
\end{equation}
where $k_z = \nu / c$  (for the counter-propagating mode we would
have $k_z = - \nu / c$, but we will concentrate here on the
co-propagating mode), $\lambda(\tau) = \mu E'(\tau) / \hbar$,
$\mu$ is the atomic dipole moment and $E'(\tau) = E \exp(-\alpha
\tau)$ is the electric field in the IRF. The spacetime coordinates
$t(\tau)$ and $x(\tau)$ for the trajectory of the accelerated
atom are given by equations~(\ref{accel}) with a redefined origin
of proper time, i.e.,  $\tau = \tau' - \tau_\mathrm{i}$, such
that the injection time corresponds to $\tau = 0$; in those
coordinates the cavity injection point is located at $x(0) = c /
\alpha$.

The coarse-grained evolution equation for a large number of atoms
injected at random times with an average injection rate (number of
atoms per unit of time) $r$ is governed by the following
Pauli-type master equation:
\begin{equation}
\fl \frac{\rmd\rho_{n,n}}{\rmd t} = -R_2 \left[(n+1) \rho_{n,n} -
n \rho_{n-1,n-1} \right] -R_1 \left[n \rho_{n,n} - (n+1)
\rho_{n+1,n+1}\right] \label{pauli} ,
\end{equation}
where $n$ is the number of photons in the cavity and $R_1$ and
$R_2$ are, respectively, the absorption and emission coefficients
computed for an empty cavity, whose precise form is given below.
The absorption and stimulated emission when there are $n$ photons
in the cavity is already accounted by the appropriate $n$ and
$(n+1)$ factors in equation~(\ref{pauli}). If $R_1 > R_2$, there
is a stationary solution which corresponds to a thermal density
matrix at temperature $\mathcal{T}_c = (\hbar \nu / k_\mathrm{B})
\ln (R_1/R_2)$.

From equation~(\ref{single}) for a single injected atom and
taking into account that the injection rate is $r$, the
transition coefficients $R_{1,2}$ for an empty cavity are
\begin{equation}
R_{1,2} = r \left| \frac{1}{\hbar} \int_{0}^{T} V_{1,2} (\tau)
\right|^2 \rmd\tau ,
\end{equation}
 with $V_1 (\tau)
= \langle e,0| \hat{H}_\mathrm{int}(\tau) |g,1\rangle$ and $V_2
(\tau) = \langle e,1| \hat{H}_\mathrm{int}(\tau) |g,0\rangle$,
where $|g\rangle$ and $|e\rangle$ denote the ground state and the
excited state of the atom respectively. Introducing the constant
$\lambda$ such that $\lambda(\tau) = \lambda \exp(-\alpha \tau)$,
the transition coefficients can be written as $R_{1,2} = r
\lambda^2 |I_{1,2} (\omega) |^2$, with $I_2 (\omega) = I_1
(-\omega)$ and $I_1 (\omega)$ given by
\begin{equation}
I_{1}(\omega) = \left(\int^T_{-\infty} \rmd\tau -
\int^0_{-\infty} \rmd\tau \right) \exp \left(i \frac{\nu}{\alpha}
\mathrm{e}^{-\alpha\tau} + i \omega\tau - \alpha\tau \right)
\label{I1},
\end{equation}
where we decomposed the integral $\int^T_{0} d\tau$ into two
separate parts for later convenience and some adiabatic switch on
at $\tau = -\infty$ (or an equivalent analytic continuation) may
be required to get a finite result. The emission coefficient
corresponds to the so-called counter-rotating terms that would
have been discarded in the rotating wave approximation~(RWA).

If we just consider the first integral in equation~(\ref{I1}), we
get an exponentially suppressed ratio for the transition
coefficients: $R_2 / R_1 = \exp (-2 \pi \omega / \alpha)$. This
corresponds to the case in which the atom-field interaction has
been present for a long time in the past. On the other hand, when
the second integral in equation~(\ref{I1}) is also included,
which corresponds to switching on the interaction instantaneously
at $\tau = 0$ (the time at which the atom is injected into the
cavity), the situation changes drastically. In particular, if we
consider the regime $\nu \gg \omega \gg \alpha$, the ratio
becomes $R_2/R_1 \simeq \alpha/(2\pi\omega)$, which implies an
enhancement of many orders of magnitude for realistic values of
$\nu$, $\omega$ and $\alpha$~\footnote{From
equations~(\ref{accel}) it follows straightforwardly that the
conditions $\exp(-\alpha\tau) \ll 1$ and $\nu \gg \alpha$ imply
that the size of the cavity should be much larger than the
wavelength of the cavity mode.}. It is this enhancement factor
which Scully {\it et al} claim to give a better sensitivity in the
detection of the radiation which they regard as due to the
accelerated motion. We will see that this process is not
responsible for the Unruh effect.

\subsection{Discussion: This scheme does not detect Unruh effect}

We focus here just on two factors, that related to the presence
of cavity, and the effect of sudden switching on of the
interaction.

\paragraph{Effect of cavity} The Unruh effect corresponds to the fact that a uniformly
accelerated particle detector (the two-level atom in this case)
perceives the vacuum fluctuations as a thermal bath. The presence
of the cavity is going to change that situation because, in
contrast to free space, the mode spectrum of the electromagnetic
fields inside the cavity is no longer Lorentz invariant.
Therefore, the effect of the vacuum fluctuations on the
accelerated atom (characterized by the noise and dissipation
kernels introduced above) will not be stationary and cannot
correspond to a thermal bath.

We emphasize again that, strictly speaking, the Unruh effect
refers to the thermal bath perceived by the detector, not to
possible radiation emitted by the detector from the point of view
of inertial observers. In fact, as discussed in the prior section,
in free space (and under some idealized conditions), there is no
real radiation emitted by the accelerated detector, i.e. no energy
flux, just a modification of the vacuum polarization.

In the scheme of Scully {\it et al}~\cite{Scully03}, briefly
summarized in the previous subsection, there is some probability
for the cavity mode to become excited when an atom is accelerated
inside the cavity. If the atom-field interaction is somehow
switched on adiabatically, the ratio of the emission and
absorption coefficients is exponentially suppressed by the
Boltzmann factor for a temperature $\mathcal{T}_c = \hbar \alpha
/ (2\pi k_\mathrm{B})$, which coincides with the temperature of
the thermal bath perceived by a UAD in free space with the same
acceleration. The reason for such a coincidence can be understood
qualitatively as follows: in the ``golden rule'' limit (limit of
large $T$ with finite $\lambda ^2 T$) one can show that the ratio
of excitation and de-excitation of a two-level detector with
characteristic frequency $\omega$ induced by each inertial mode
in free space is given by the same Boltzmann factor $\exp(- 2 \pi
\omega / \alpha)$. Nevertheless, the thermal distribution of
photons in the cavity is not in one-to-one correspondence with
the Unruh effect for several reasons. First, the atoms
accelerated inside the cavity are not in thermal equilibrium
because they do not perceive the vacuum fluctuations in the
cavity as a thermal bath, as explained above. Second, the thermal
population of photons in the cavity results from a statistically
independent events of injecting a sufficient number of atoms at
random times. Third, an analogous thermal population can be
obtained even by injecting atoms with constant velocity, because
the sudden switching on of interaction can produce such a result,
as described below.

\paragraph{Effect of sudden switch-on}  The fact that the atoms are injected
into the cavity at some initial time is effectively equivalent to
suddenly switching on the atom-field interaction. In that case,
the ratio of emission and absorption is enhanced. In particular,
in the regime $\nu \gg \omega \gg \alpha$, it is given by
$R_2/R_1 \simeq \alpha/(2\pi\omega)$. As recognized in
reference~\cite{Scully03}, this is entirely due to the
nonadiabatic switching on of the interaction. What is more
relevant to the issues at hand is that with non-adiabatic
switch-on, the acceleration no longer plays a crucial role.
Indeed, in that regime the emission rate is $\lambda ^2 |I_2|^2
\simeq \lambda ^2 / \nu^2$ and is thus independent of the
acceleration. It is true that the absorption coefficient still
depends on the acceleration, which is a consequence of the
resonance when the increasingly redshifted frequency $\nu \exp(-
\alpha \tau)$ of the cavity mode as perceived by the atom
coincides with the frequency $\omega$ of the atom, but this is
not essential.

This important point can be illustrated by considering the case in
which the atoms are injected with constant velocity into the
cavity. As mentioned in reference~\cite{Scully03}, the ratio
becomes then
\begin{equation}
\frac{R_2}{R_1} = \left|\frac{\nu' - \omega}{\nu' + \omega}
\right|^2 \left| \frac{1-\rme^{-\rmi(\nu' + \omega)T}} {1 -
\rme^{-\rmi(\nu^{\prime} -\omega)T}} \right|^2,
\end{equation}
where $T$ is the time that the atom spends in the cavity and
$\nu' = \nu (1- v/c)^{1/2} / (1+ v/c)^{1/2}$ is the Doppler
shifted frequency of the co-propagating cavity mode (the velocity
sign should be changed for the anti-propagating mode).

Due to space restriction we can only raise two issues in this
detection scheme. Even though this detection scheme, like a few
earlier ones proposed, does not capture the essence of Unruh
effect, it is still a good intellectual exercise to expound the
physics of the core processes in the proposals.  This is because
Unruh effect brings up interesting new issues related to the
vacuum fluctuations, polarizations and energy flux in the quantum
field, as influenced by different motional states of the atoms or
detectors.

\paragraph{Acknowledgments}
BLH gladly acknowledges Alpan Raval, Don Koks, Mei-Ling Tseng and
Phil Johnson for earlier fruitful collaborations on the problems
of emitted radiation, vacuum polarization and radiation damping
from moving detectors and charges, which form the basis of our
current investigation. He thanks Steve Fulling for introducing
(enlisting) him  to think about the Scully {\it et al} proposal.
He also thanks Paul Davies and Gerard Milburn for mentioning
Pendry's work and their ideas on dropping atoms to measure vacuum
friction. AR and BLH thank Stefano Liberati for discussions on
features of this detection scheme and Luis Orozco, Bill Phillips
and Steve Rolston for useful comments. This work is supported in
part by grants from NSF PHY03-00710, NIST and contract from
ARDA-LPS.

\section*{References}


\end{document}